\documentclass[twocolumn]{aastex631}
\usepackage{amsfonts}
\usepackage{amsmath}
\usepackage{amssymb}
\usepackage{xcolor}
\usepackage{graphicx}
\usepackage{subfigure}
\usepackage{slashed}
\usepackage{hyperref}
\usepackage{amsthm}
\usepackage{mathrsfs}
\usepackage{color}
\usepackage{epsfig}
\usepackage{epstopdf}
\usepackage{bm}
\usepackage[T1]{fontenc}
\usepackage{ae,aecompl}	
\usepackage[toc,page]{appendix}
\usepackage{soul}
\usepackage[utf8]{inputenc}
\usepackage{multirow}
\usepackage{verbatim}

\def\be{\begin{eqnarray}}\def\ee{\end{eqnarray}}

\def\lsim{\mathrel{\rlap{\lower3pt\hbox{\hskip1pt$\sim$}}
		\raise1pt\hbox{$<$}}} 
\def\gsim{\mathrel{\rlap{\lower3pt\hbox{\hskip1pt$\sim$}}
		\raise1pt\hbox{$>$}}} 

\newcommand\sect[1]{\section{#1}}

\begin{document}

\title{Insights into neutron star equation of state by machine learning}

\author{Ling-Jun Guo}
\affiliation{College of Physics, Jilin University, Changchun, 130012, China}
\affiliation{School of Fundamental Physics and Mathematical Sciences, Hangzhou Institute for Advanced Study, UCAS, Hangzhou, 310024, China}

\author{Jia-Ying Xiong}
\affiliation{School of Fundamental Physics and Mathematical Sciences, Hangzhou Institute for Advanced Study, UCAS, Hangzhou, 310024, China}
\affiliation{Institute of Theoretical Physics, Chinese Academy of Sciences, Beijing 100190, China}

\author{Yao Ma}
\email{mayao@ucas.ac.cn}
\affiliation{School of Fundamental Physics and Mathematical Sciences, Hangzhou Institute for Advanced Study, UCAS, Hangzhou, 310024, China}
\affiliation{University of Chinese Academy of Sciences, Beiing 100049, China}
\affiliation{TaiJi Laboratory for Gravitational Wave Universe (Beijing/Hangzhou), University of Chinese Academy of Sciences, Beijing, 100049, China}

\author{Yong-Liang Ma}
\email{ylma@nju.edu.cn}
\affiliation{Nanjing University, Suzhou, 215163, China}
\affiliation{School of Fundamental Physics and Mathematical Sciences,
	Hangzhou Institute for Advanced Study, UCAS, Hangzhou, 310024, China}
\affiliation{TaiJi Laboratory for Gravitational Wave Universe (Beijing/Hangzhou), University of Chinese Academy of Sciences, Beijing, 100049, China}
\affiliation{International Center for Theoretical Physics Asia-Pacific (ICTP-AP) , UCAS, Beijing, 100190, China}

\date{\today}

\begin{abstract}
	Due to its powerful capability and high efficiency in big data analysis, machine learning has been applied in various fields. We construct a neural network platform to constrain the behaviors of the equation of state of nuclear matter with respect to the properties of nuclear matter at saturation density and the properties of neutron stars. It is found that the neural network is able to give reasonable predictions of parameter space and provide new hints into the constraints of hadron interactions. As a specific example, we take the relativistic mean field approximation in a widely accepted Walecka-type model to illustrate the feasibility and efficiency of the platform. The results show that the neural network can indeed estimate the parameters of the model at a certain precision such that both the properties of nuclear matter around saturation density and global properties of neutron stars can be saturated. The optimization of the present modularly designed neural network and extension to other effective models is straightforward.
\end{abstract}


\sect{Introduction}
With the development of technologies, the amount of experimental data is increasing rapidly. These massive data are usually from different experimental targets and provide information on different aspects of the same system. Therefore, in order to get a complete understanding of the physics involved, one needs to combine these data and analyze them systematically. However, it is usually a challenge for researchers to handle this kind of process due to the complexity of the data and the massive parameters in the model. The recently developed machine-learning (ML) or artificial intelligence(AI)-driven technologies provide a way out. The ML methods have already earned credit in the fields of big data analysis due to their advantages of efficiency and adaptivity~\cite{krizhevsky2012imagenet,JorSc349,LeCNt521,li2019deep}, and they have already been applied in many different fields of physics, e.g., Refs.~\cite{abraham2018detection,Mehta:2018dln,niu2019deep,gu2020machine,Brady:2021plj,Boehnlein:2021eym,He:2023zin,He:2023urp,oala2023data} and references therein.

In nuclear physics, the properties of nuclear matter (NM) have been investigated for a long period, but no consensus has been arrived at. Some fundamental questions are waiting for clarification, such as what the components of the NM are, whether there is a phase transition in the dense matter of compact stars, and whether dense NM exists in states other than Fermi liquid, among others (see, e.g., reviews~\cite{Brown:2001nh,Hayano:2008vn,Holt:2014hma,Baym:2017whm,Li:2019xxz,Ma:2019ery,Burgio:2021vgk,Ma:2020nih,Ma:2023ugl} and references therein). To resolve these questions, all the existing information from both experiments and theories should be combined in the corresponding analysis; then a reliable technique, like ML developed here, is necessary. 

The constraints on NM come from both terrestrial experiments and astrophysical observations. Owing to the analysis of the structures of heavy nuclei, e.g., ${ }^{24} \mathrm{Mg}, { }^{90} \mathrm{Zr}, { }^{116} \mathrm{Sn}$, and ${ }^{208} \mathrm{~Pb}$, and the data from heavy-ion collisions, one can provide information about NM properties around nuclear saturation density ($n_0 \approx 0.16 \mathrm{fm}^{-3}$;~\cite{youngblood1999incompressibility,brown2000neutron,karataglidis2002discerning,steiner2005isospin}), such as the binding energy of nucleon $e_0$, the symmetry energy $E_{\mathrm{sym}}$, the incompressibility coefficient $K_0$, the skewness coefficient $L_0$, and so on. Besides the above information obtained from terrestrial experiments, there are also constraints from astrophysics. The signals from, e.g., PSR J1614-2230, J0348+043, PSR J0740+6620, J0030+0451 and PSR J0740+662 constrain the mass-radius (MR) relation of neutron stars (NSs;~\cite{Demorest:2010bx,Antoniadis:2013pzd,Ozel:2016oaf,Fonseca:2016tux,NANOGrav:2019jur,Fonseca:2021wxt}). The observation of gravitational waves (GWs) from the binary NS merger, GW170817~\cite{LIGOScientific:2017vwq,LIGOScientific:2018cki}, yields a new but more rigorous independent constraint---tidal deformation---on the MR relations~\cite{Annala:2017llu,Annala:2021gom} and the multimessenger era of NSs has begun~\cite{hartley2017multi,icecube2018multimessenger}. Therefore, in the study on NM, both the data from terrestrial experiments and astrophysical observations should be considered, and these massive data increase the difficulty of the analysis process.

On the other hand, the equation of state (EoS) plays a key role in the studies on NM, and it is usually parameterized by effective models or theories, e.g., one-boson-exchange (OBE) model~\cite{Machleidt:1987hj} and chiral effective field theory ($\chi {\mathrm{EFT}}$; ~\cite{Epelbaum:2008ga,Ma:2019kiq}), due to the nonperturbative nature of quantum chromodynamics (QCD). The information of QCD is encoded into the low-energy constants (LECs) of these models and theories, which are determined by fitting the experimental data, but the fitting process is somehow fine-tuning the values of the LECs because of the cancellation mechanism between attractive and repulsive terms. One can easily envision that this type of parameter space exploration can be very laborious and time intensive.

In practice, the simultaneous treatment of the massive data and fine-tuning of the LECs to obtain a physical EoS is very difficult if not unfeasible. With the help of ML methods, the cost of the labor- and time-consuming work can be reduced. The recently developed AI technologies~\cite{7298965,10.5555/3295222.3295349}, can simplify these process significantly, and some remarkable work has already been done along this line; see, e.g.,~\cite{drischler2020well, Fujimoto:2019hxv,Morawski:2020izm,Traversi:2020dho, Fujimoto:2021zas, Morawski:2020izm,ferreira2021unveiling,Annala:2021gom,Krastev:2021reh,Ferreira:2022nwh,Soma:2022qnv,Chatterjee:2023ecc,Zhou:2023cfs,Krastev:2023fnh}, where several parameterizations of NM EoS, e.g. SLy4, APR, and BSK20, are used with the help of neural network (NN) to understand NM properties and NS structures.

The purpose of this work is to propose an ML platform to carry out the above idea. We will construct an NN with respect to the properties of NM around saturation density and the MR relations of NSs to constrain the behaviors of EoS. A specific but widely accepted nucleon force model is adopted as a preliminary illustration of the efficiency of this NN. This platform can be easily extended to other complex, generalizable, and practical models.

\section{Neural network framework and its application}
\label{sec:NN}

\begin{figure*}[htbp]
			\centering
			\includegraphics[scale=0.2]{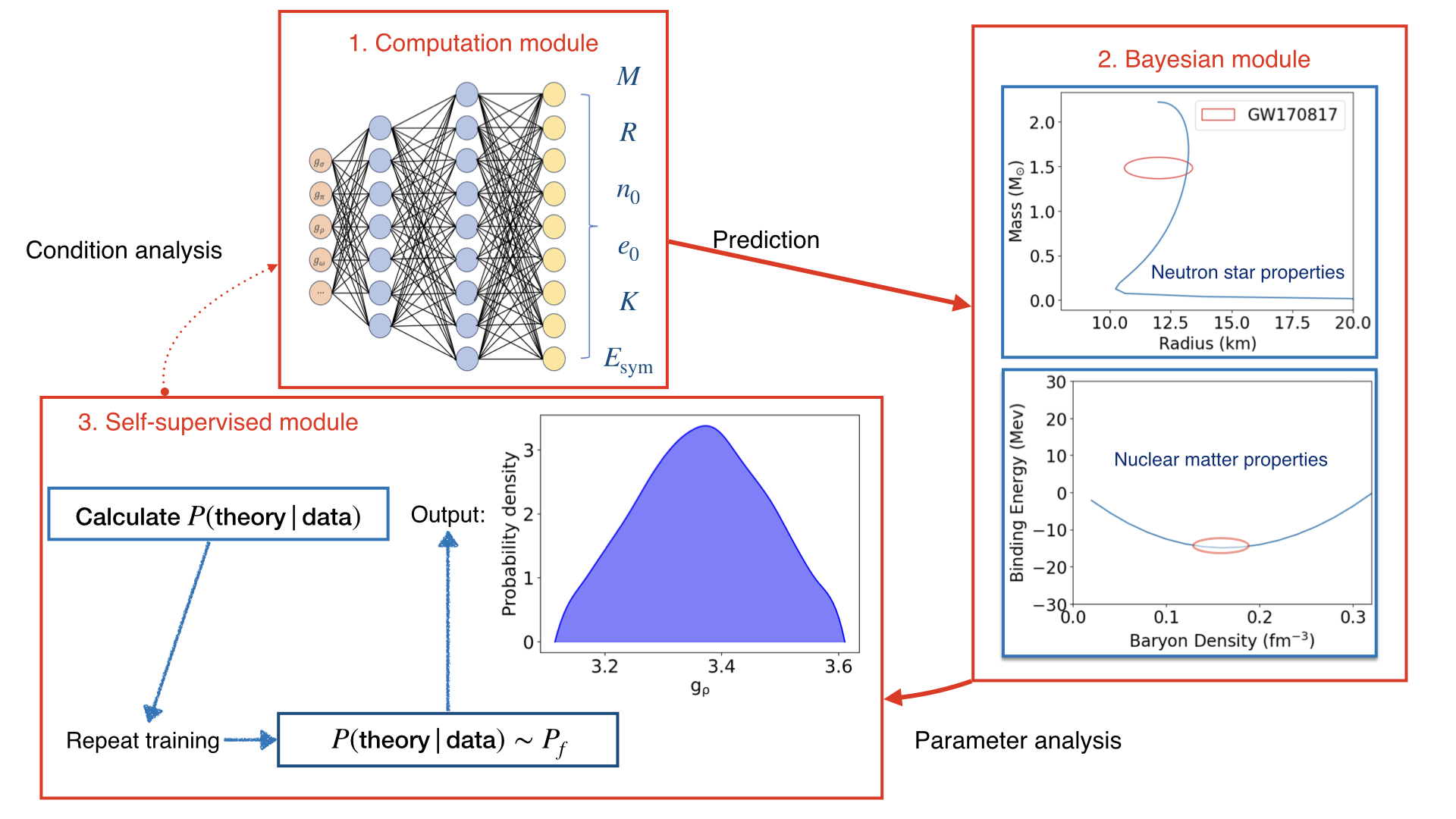}
			\caption{Structure of NN platform.}
			\label{fig:NNf}
		\end{figure*}

The structure of our NN platform for the parameter-searching processes is shown in Figure~\ref{fig:NNf}.
Concretely, the strategy of the NN is the following: At first, a model or theory with its prechosen parameter space should be given to train the computation module of NN, and it does not matter whether the prechosen parameter space is a physical one or not.
Then, the trained computation module predicts possible values of parameters according to the Bayesian module calculation by experimental data from NM and NS. If the self-supervised module judges that the data set indicates a parameter set beyond the training set, it will make possible predictions and generate a corresponding training set. In this situation, the new training set and the previous one will be combined by an adaptive algorithm to train the computation module again, and this process will repeat until the probability from Bayesian analysis converges to a fixed value, which represents the confidence level of the input model.

\subsection{Application in a specific model}

After a discussion on the basic idea of the NN, we illustrate its application in this part by using a widely used nuclear force model~\cite{Sugahara:1993wz,Shen:1998gq}. The model takes the following form:
\be
\mathcal{L}_{\mathrm{RMF}} & = & \bar{\psi}\left[i \gamma_\mu \partial^\mu-M-g_\sigma \sigma-g_\omega \gamma_\mu \omega^\mu-g_\rho \gamma_\mu \tau_a \rho^{a \mu}\right] \psi \nonumber\\
& &{} +\frac{1}{2} \partial_\mu \sigma \partial^\mu \sigma-\frac{1}{2} m_\sigma^2 \sigma^2-\frac{1}{3} g_2 \sigma^3-\frac{1}{4} g_3 \sigma^4 \nonumber\\
& &{} -\frac{1}{4} W_{\mu \nu} W^{\mu \nu}+\frac{1}{2} m_\omega^2 \omega_\mu \omega^\mu+\frac{1}{4} c_3\left(\omega_\mu \omega^\mu\right)^2 \nonumber\\
& &{} -\frac{1}{4} R_{\mu \nu}^a R^{a \mu \nu}+\frac{1}{2} m_\rho^2 \rho_\mu^a \rho^{a \mu},
\label{eq:Lag}
\ee
where $\psi$ is the isodoublet of nucleon field with mass $M$; $\sigma$ is the isoscalar scalar meson field; and, $\omega^\mu$ and $\rho^{\mu}$ are, respectively, the isoscalar and isovector meson fields with $W^{\mu \nu}$ and $R^{a\mu \nu}$ being their field strength tensors:
\be
W^{\mu \nu} & = & \partial^\mu \omega^\nu-\partial^\nu \omega^\mu\ , \nonumber\\
R^{a \mu\nu} & = & \partial^\mu \rho^{a \nu}-\partial^\nu \rho^{a \mu}+g_\rho \epsilon^{a b c} \rho^{b \mu} \rho^{c \nu}\ .
\ee
In this model, the nucleon-meson coupling terms are the most simple OBE-type interactions, the nonlinear $\sigma$ terms are crucial to describe incompressibility, and the nonlinear $\omega$ terms are added in order to reproduce the density dependence of nucleon self-energy.

In practice, to calculate the NM properties using Model~(\ref{eq:Lag}), some approximations should be applied. In the relativistic mean field (RMF) approximation, which is widely used, the problem is reduced to solve the following coupled equations of motion (EoMs):
\be
m_{\sigma}^2\sigma+g_2\sigma^2+g_3\sigma^3 & = & {} -g_{\sigma}\left(\rho_{n,s}+\rho_{p,s}\right)\ ,\nonumber\\
m_{\omega}^2\omega+c_3\omega^3 & = & g_{\omega}\left(\rho_p+\rho_n\right)\ ,\nonumber\\
m_{\rho}^2\rho & = & g_{\rho}\left(\rho_p-\rho_n\right)\ ,
\label{eq:rmfeom}
\ee
where $\rho_{n(p)}$ and $\rho_{n,s(p,s)}$ are, respectively, the density and scalar density of neutron (proton) with
\be
\rho_{n(p),s} & = & \frac{m_N^{*3}}{\pi^2}\int_{0}^{t_{n(p)}}{\rm{d}}x\frac{x^2}{\sqrt{1+x^2}} \nonumber\\
& = & \frac{m_N^{*3}}{\pi^2}\left[\frac{1}{2}\left(t_{n(p)}\sqrt{1+t_{n(p)}^2}-{\rm{arcsinh}}{t_{n(p)}}\right)\right]\ ,
\ee
with $m_N^{*}=M+g_{\sigma}\sigma$ being the effective mass of nucleons and $t_{n(p)}=\frac{(3\pi^2\rho_{n(p)})^{1/3}}{m_N^*}$. After determining the expectations of meson fields by solving the EoSs~(\ref{eq:rmfeom}), the energy density $\mathcal{E}$ can be obtained by calculating the Hamiltonian, and the pressure $P$ can be obtained via $P=-\mathcal{E}+\rho_{N}\frac{\rm{d}\mathcal{E}}{\rm{d}\rho_N}$.

Usually, in a model of NM, the solution of the EoMs and the prediction of the EoS suffer from the constraints from the NM properties around saturation density and the mathematical structure of the EoMs. From the mathematical structure of the EoMs~(\ref{eq:rmfeom}), it can be seen that even in the simple Model~(\ref{eq:Lag}), the EoMs have the multiroots problem, which seemingly lead to the NN convergence problem. But one can expect the results given by the NN are smooth and physical due to the mathematical reasons: (i) identity theorem of analytic functions and (ii) the full rank properties of a equation group coefficient matrix. Therefore, the solution planes of Equation~(\ref{eq:rmfeom}) will only intersect or be tangent if the natural condition is assumed. In either case, the solution plane given by NN will be defined uniquely by choosing a starting point, which will be the origin point in field expectation solution space in the following discussions.

\begin{figure*}[htbp]
	\centering
	\includegraphics[scale=0.13]{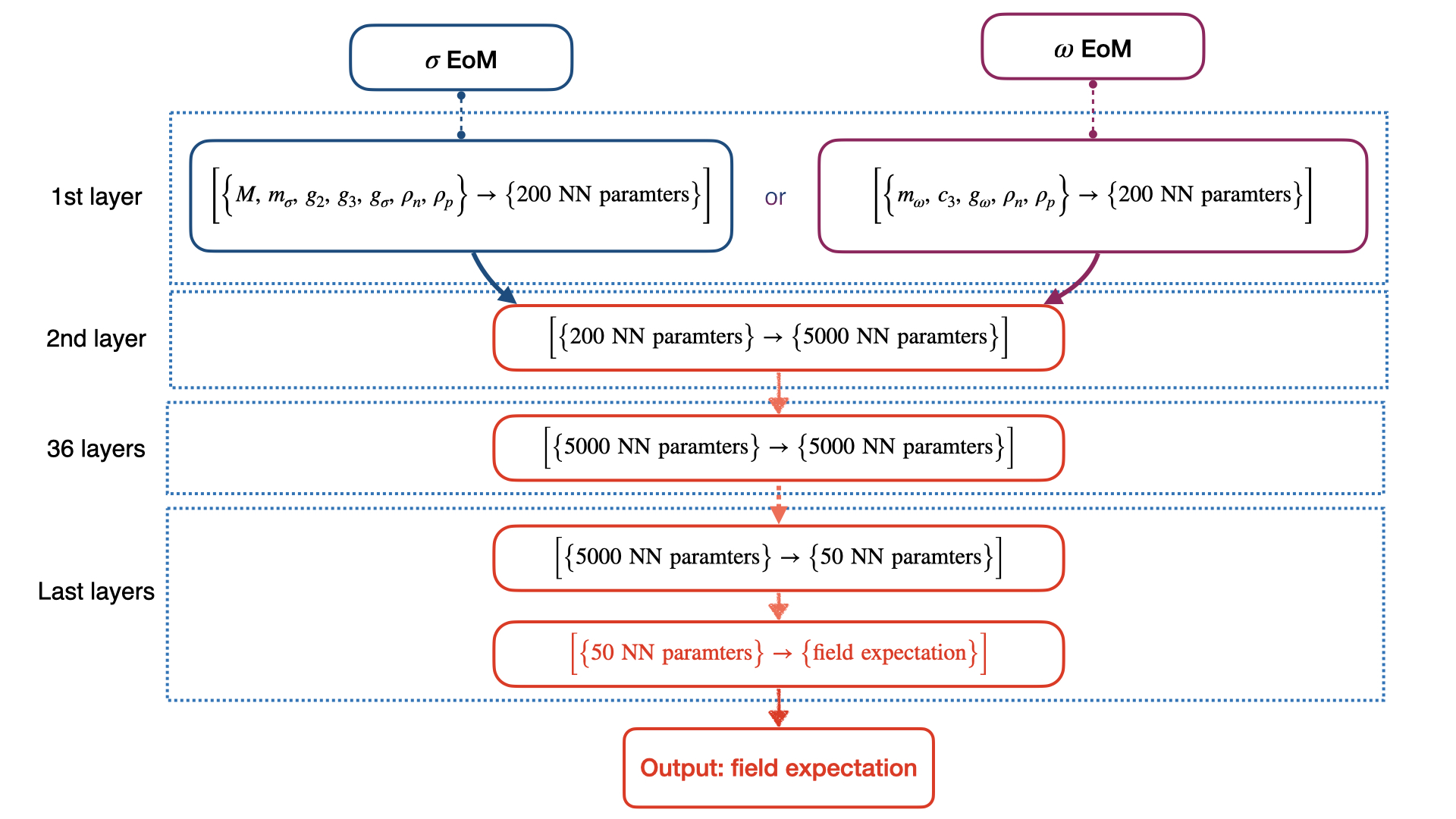}
	\caption{Layers to solve EoMs of meson fields. ReLU is used as activation function~\cite{Glorot2011DeepSR}. The layers' construction for $\rho$ field is ignored in current work for simplicity since its EoM in Equation~(\ref{eq:rmfeom}) is just a linear function of nucleon density.}
	\label{fig:Layers}
\end{figure*}

In the explicit calculation, the computation module of NN~\footnote{The NN in this work is built on PyTorch platform.~\cite{10.5555/3454287.3455008}} to solve the above equation group is based on 40 fully connected layers, where there is a batch normalization layer between each set of two fully connected layers to accelerate the training~\cite{10.5555/3045118.3045167}.
The fully connected layers are constructed as Figure~\ref{fig:Layers}.
The loss function in the above NN is defined as 
\be
f_{\rm{loss},\sigma} & = & \left[m_{\sigma}^2\sigma+g_2\sigma^2+g_3\sigma^3 + g_{\sigma}\left(\rho_{n,s}+\rho_{p,s}\right)\right]^2\ ,\nonumber\\
f_{\rm{loss},\omega} & = & \left[m_{\omega}^2\omega+c_3\omega^3-g_{\omega}\left(\rho_p+\rho_n\right)\right]^2\ ,
\ee
which enhances the ability of NN to give a smooth and physical prediction. The pretraining set is generated randomly at regions where $\left\{\rho_n,\ \rho_p\right\}\in\left[\left\{0,0\right\},\left\{10n_0,10n_0\right\}\right]$ with distribution function $\mathcal{P}(\rho_n,\rho_p)=\frac{\sqrt{10}}{20}\sqrt{\frac{n_0}{\rho_n+\rho_p}}$ in order to improve the ability to describe low-density regions. The learning process is done with the help of the Adam algorithm~\cite{Kingma2014AdamAM}. Based on the results of solved EoMs, the EoS can be obtained from the Lagrangian~(equation (\ref{eq:Lag})), and star properties can be calculated by solving the Tolman-Oppenheimer-Volkoff (TOV) equation~\cite{Oppenheimer:1939ne,Tolman:1939jz}.

In the numerical calculation by using the NN constructed above, we impose the constraints from the NS properties and the NM properties around saturation density. Our choices are listed in Table \ref{tab:input} which are estimated from \cite{Colo:2004mj,Dutra:2012mb,Dutra:2014qga,LIGOScientific:2018cki,LIGOScientific:2018hze,Choi:2020eun}

		\begin{table*}[ht]
			\caption { $M$ and $ R$ is the mass and radius of the primary star in GW170817}
			\label{tab:input}
			\begin{tabular}{ccccccc}
				\hline
				\hline
				~ & $M$~($M_\odot$) & $R$~(km) & $n_0$~(fm$^{-3}$) & $e_0$~(MeV) & $E_{\rm sym}$~(MeV) & $K_0$~(MeV)  \cr
				\hline
				Constraints & ~$1.48 \pm 0.12 $ &~ $11.9 \pm 1.4 $ &~ $0.155 \pm 0.05$ & $-16\pm 1.0$ & ~$32.0 \pm 2.0$ & ~$250\pm 50$ \cr
				\hline
				Optimal & ~\rule{0.5cm}{1pt} &~ \rule{0.5cm}{1pt} &~ $0.161$ & $-15.7$ & ~$28.6$ & ~$266$ \cr
				\hline
				\hline
			\end{tabular}
			
			{ \textbf{Note.} \(e_0\) is the binding energy of the nucleon, $E_{\mathrm{sym}}(n)=\left.\frac{1}{2} \frac{\partial^2 E(n, \alpha)}{\partial \alpha^2}\right|_{\alpha=0}$ is the symmetry energy, and \(K_0 = n_0 \frac{\partial^2 \varepsilon (n,0)}{\partial n^2} \bigg|_{n=n_0}\) is the incompressibility coefficient at \(n_0\), respectively. Normal distribution is assumed for data error bars with 95\% confidence level.}

		\end{table*}

\begin{figure}[htbp]
	\centering
	\includegraphics[scale=0.4]{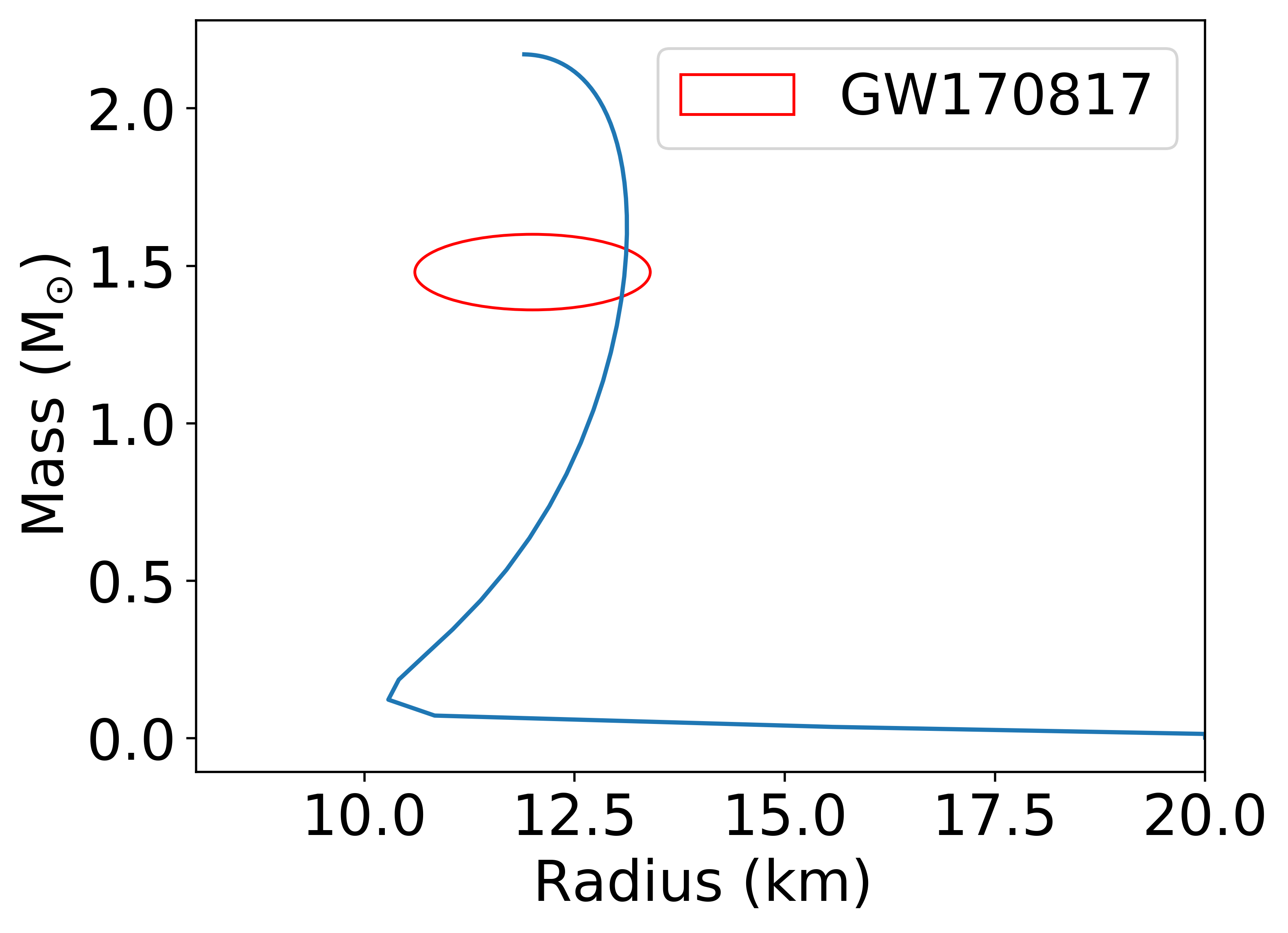}
	\caption{MR relation obtained by the optimal solution.}
	\label{fig:MRoptimal}
\end{figure}

With respect to the constraints shown in Table~\ref{tab:input} , we are ready to calculate the parameters of Model~(\ref{eq:Lag}) using the NN constructed in this work. Our results of the optimal solutions are shown in Table \ref{tab:optimal}~\footnote{
	The weight of parameter point is defined as $\mathcal{W}=\prod\mathcal{P}_i$, where $\mathcal{P}$ refers to probability density and $i$ represents physical quantities.}.
The optimal values of the parameter set lead to the NM properties shown in Table \ref{tab:input} and the corresponding MR relation illustrated in Figure~\ref{fig:MRoptimal}. From the NM properties and the MR relation, one can see that the optimal values from the NN are globally consistent with the empirical values and the observations from GW170817. This demonstrates the rationality of the NN approach. 

\begin{table}[htbp]
	\centering
	\caption { Optimal values of the parameters calculated from NN. The empirical values $M_N = 938$~MeV, $m_\omega = 782$~MeV and $m_\rho = 765$~MeV are physical values at vacuum taken as inputs for simplicity.}
	\label{tab:optimal}
	\begin{tabular}{ccccccc}
		\hline
		\hline
		$g_{\sigma}$ & $g_{\omega}$  & $g_{\rho}$ & $g_3$ & $c_3$ & $g_2$ & $ m_{\sigma}$   \cr
		\hline
		$9.82$ &~ $11.8 $ &~ $3.42$ & ~$1.26$ & ~$72.6$ & ~$-1550$~MeV &~ $531$~MeV \cr
		\hline
		\hline
	\end{tabular}
	
\end{table}

\begin{figure*}[htbp]
	\centering
	\subfigure[\(e_0\) with NM property constraints.]{\includegraphics[scale=0.035]{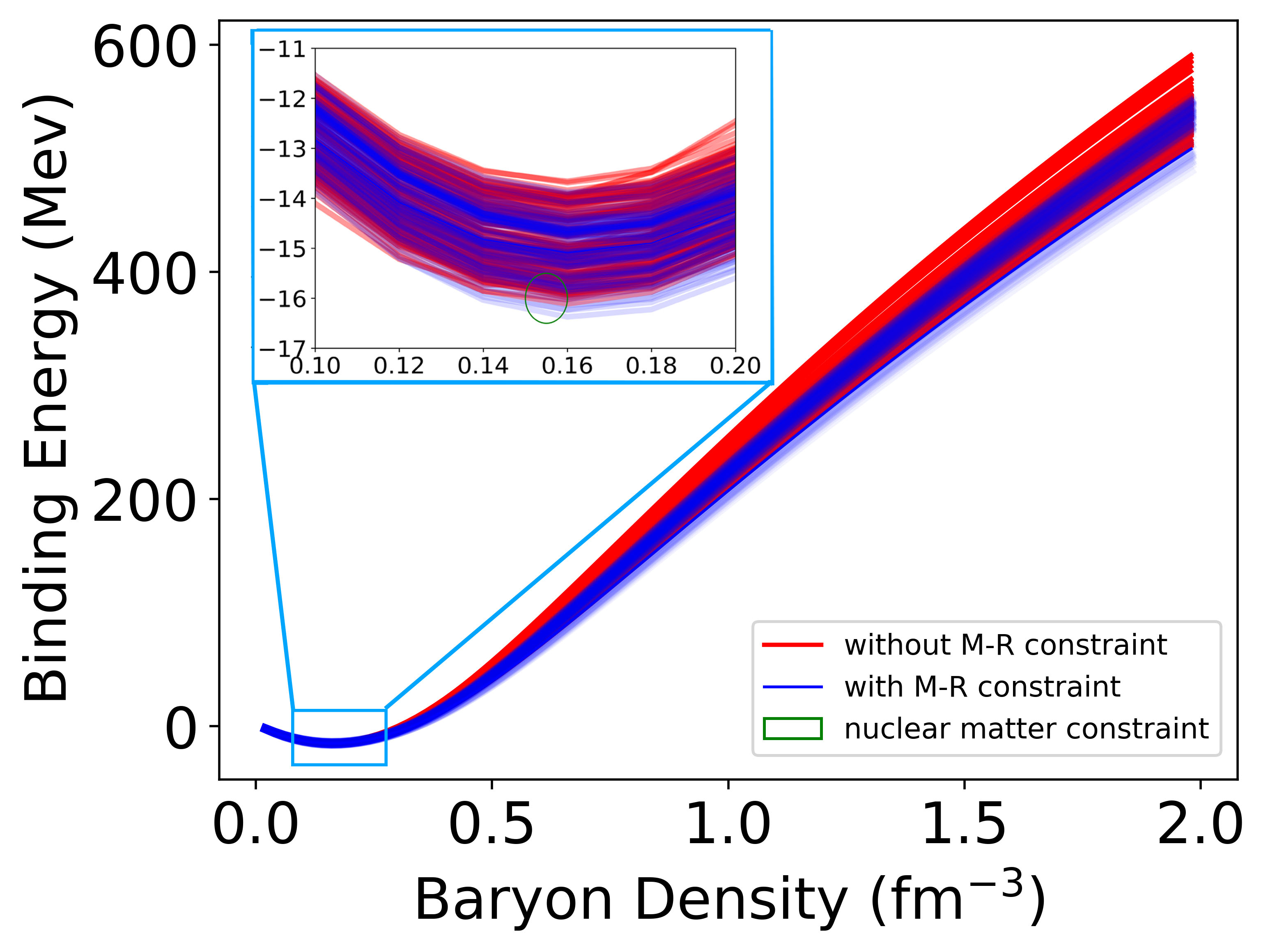}}	
	\subfigure[\(E_{\rm{sym}}\) with NM property constraints.]{\includegraphics[scale=0.035]{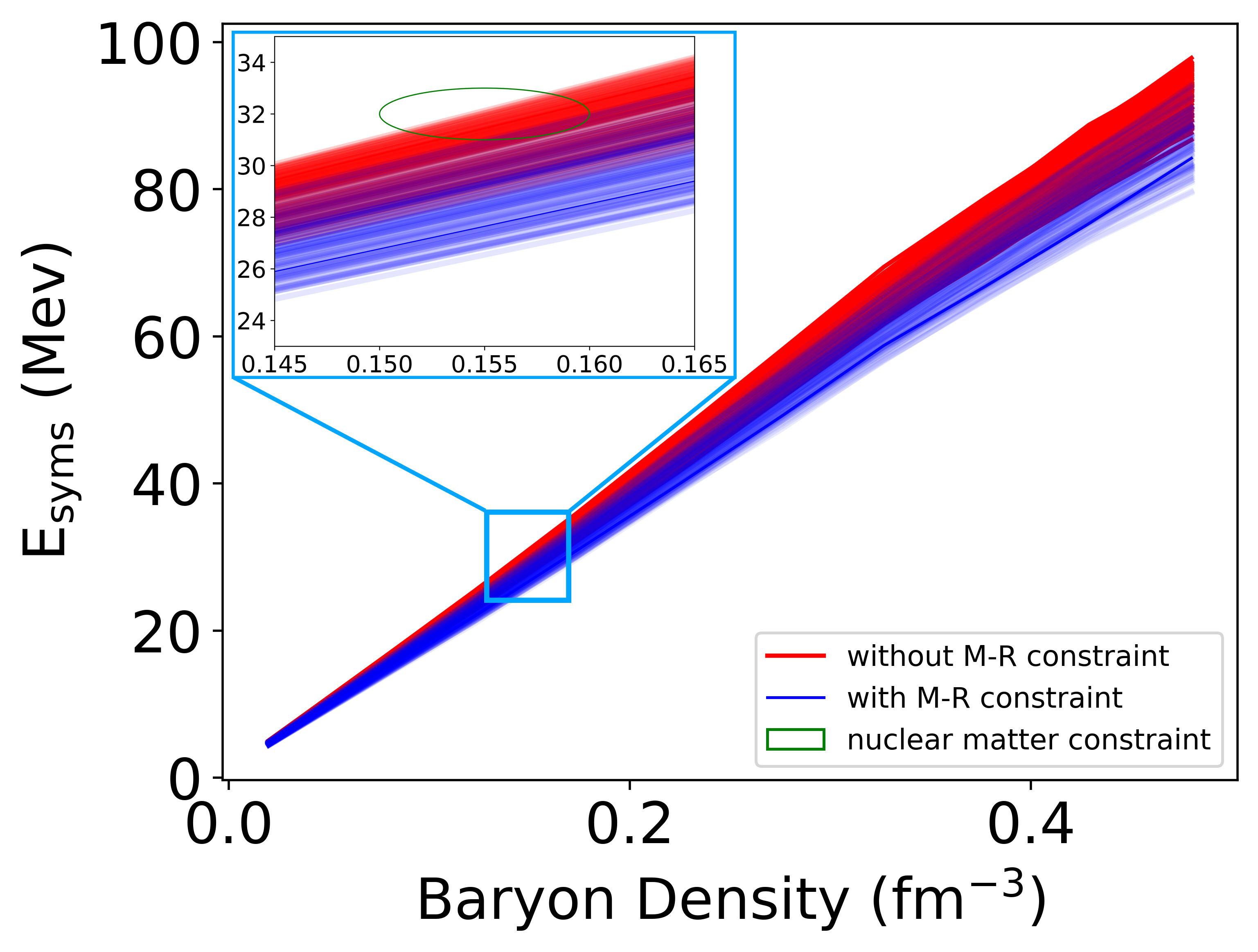}}
	\subfigure[MR relations with MR constraints.]{\includegraphics[scale=0.25]{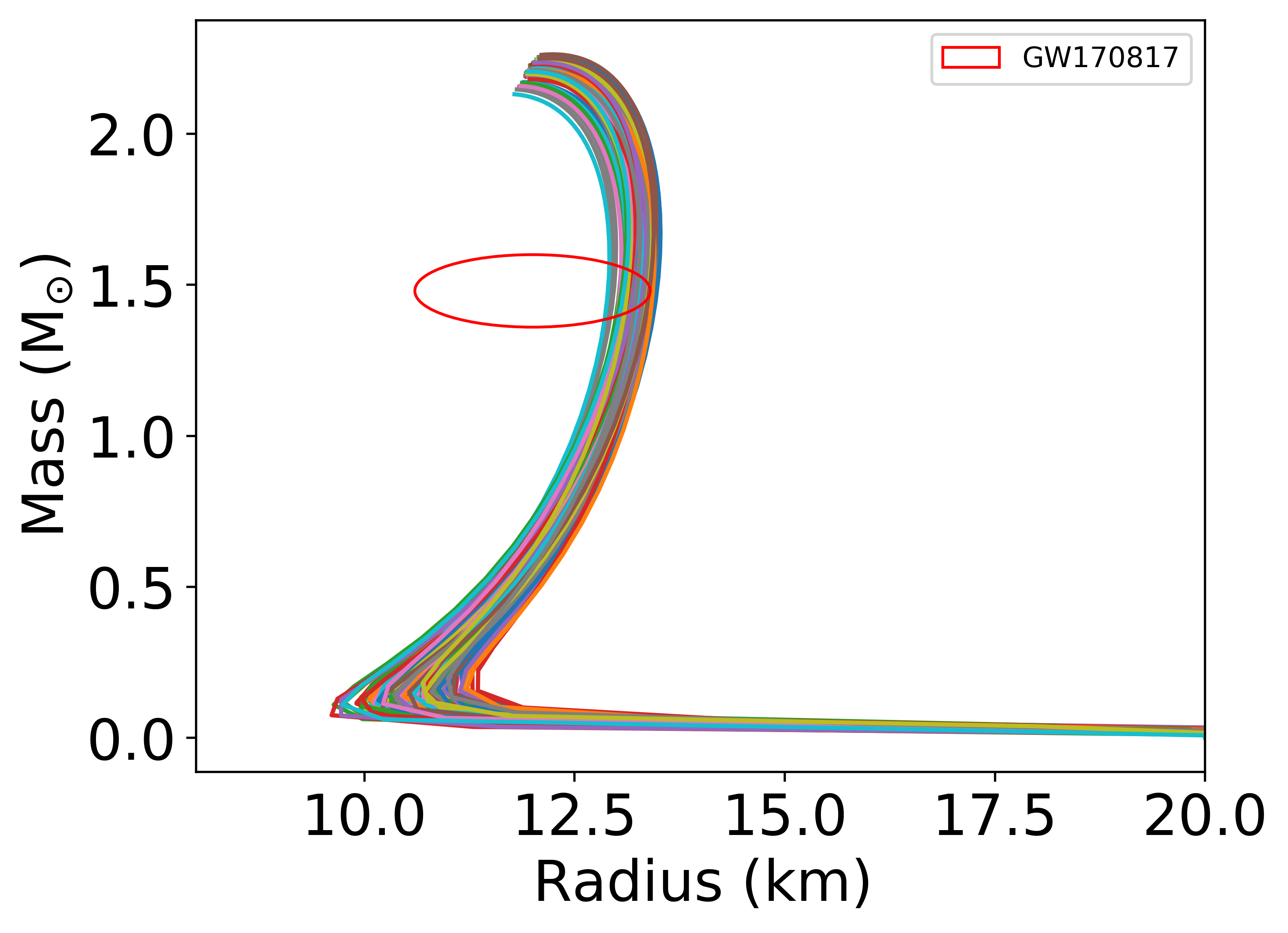}}	
	\subfigure[MR relations without MR constraints.]{\includegraphics[scale=0.25]{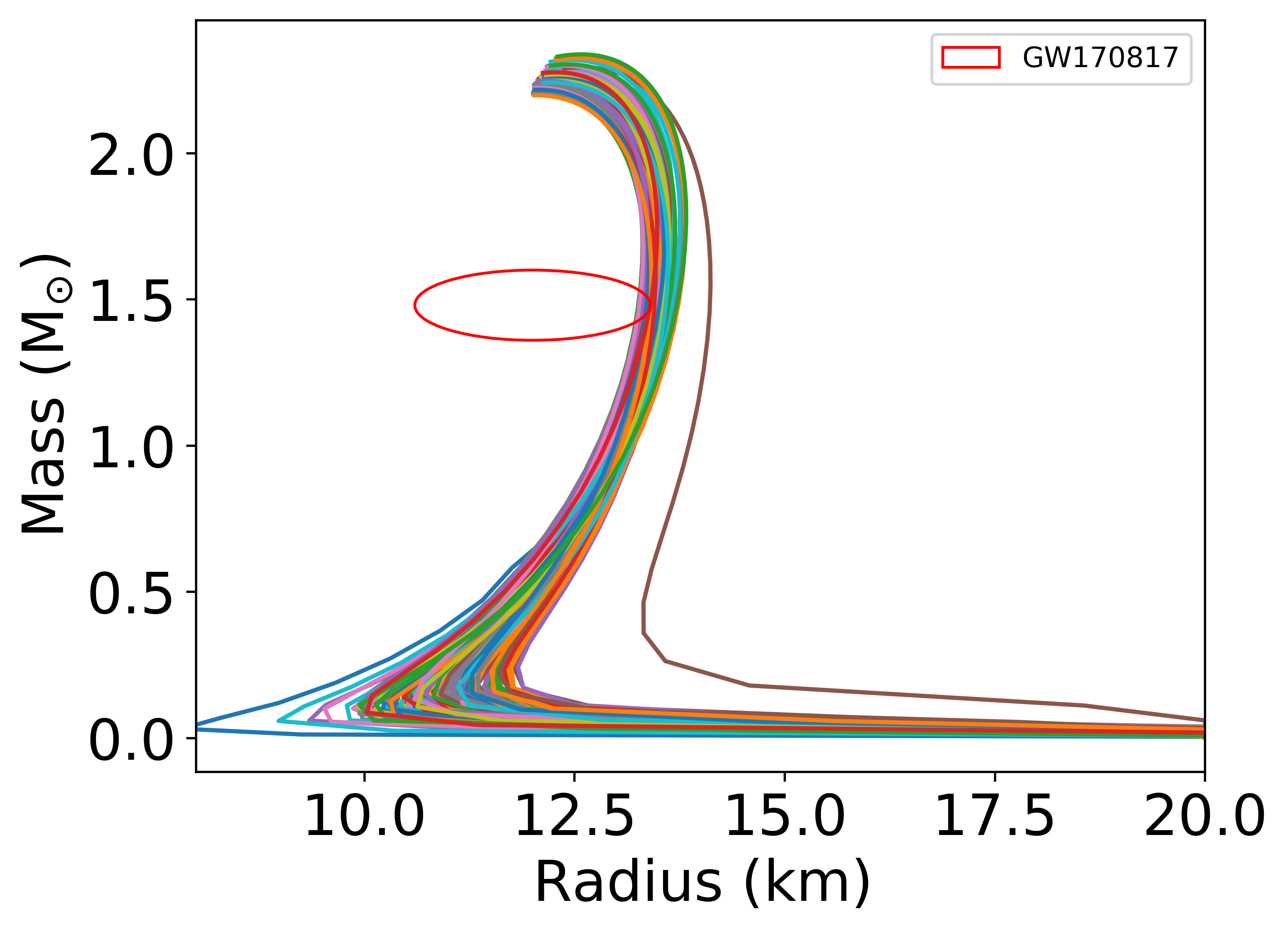}}
	\caption{(Color online) Regions of the NM properties and MR relations predicted by NN.}
	\label{fig:NN-re}
\end{figure*}

We next plot the regions of the NM properties and the MR relations from the predicted parameter space with $10.1\%$ confidence level within the data error band~\footnote{
	The confidence level is calculated by \(P_c=\int_{V}J\mathcal{W}{\mathrm{d}}V\), where the measure \(V\) is the parameter space region with \(O\) being the optimal point and \(J\) refers to the determinant of Jacobian matrix mapping data space to parameter space.
}. In Figure~\ref{fig:NN-re}, the bands of the NM properties with and without GW170817 data are presented. One can see that the inclusion of the constraints from GW170817 shrinks the bands and softens the EoS globally. This observation shows the necessity to analyze both the constraints from NS relations and NM properties simultaneously.

From Figure~\ref{fig:MRoptimal} and Table~\ref{tab:input}, one can see an interesting fact that predicted quantities by the NN with seven free parameters are not located at the center values of six constraints.
The possible reasons are: (i) the parameter space is redundant; and (ii) the degree of data freedom exceeds $7$.
The redundancy of parameter space means that at least two parameters in this system can be expressed by one parameter, which indicates the optimal solution is actually an open area, not a specific point in the parameter space.
After scanning the area around the optimal solution listed in Table~\ref{tab:optimal}, we find that it is indeed a point, which excludes the possibility of the redundancy of the parameter space.
Then the remaining reason is just that the dimension of data space is larger than seven. This is imaginable since the MR relation is obtained via the integral of EoS over the corresponding density interval, which means that NS properties are affected by the whole EoS line shape, not some specific points.
Meanwhile, from NM properties and MR relations obtained in this work shown in Figure~\ref{fig:NN-re}, it is found that the statistical significance of NM properties in the Bayesian analysis is reduced by including the constraints of the MR relation because the area obtained from the case with MR constraint deviates more from the $e_0$ and $E_{\rm sym}$ constraints but meets the MR constraint better.
It again shows the need to take care of both NM and NS properties, simultaneously, in nuclear force study.
In addition, it is found that MR lines only lie on the right side of the constraint in Figure~\ref{fig:NN-re}, which provides evidence of the NN's ability to identify data-favored models.

\begin{figure*}[htpb]
	\centering
	\subfigure{\includegraphics[scale=0.25]{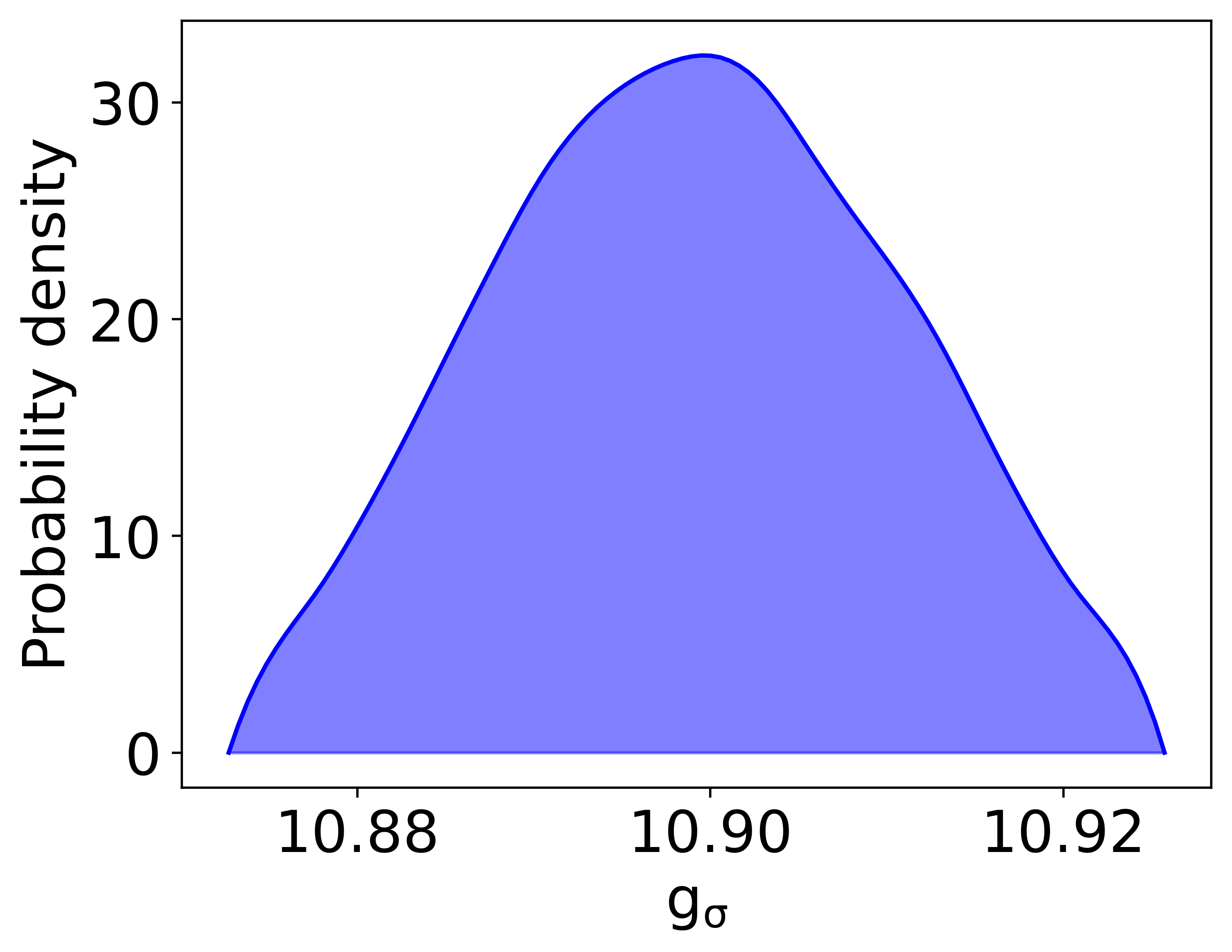}}
	\subfigure{\includegraphics[scale=0.25]{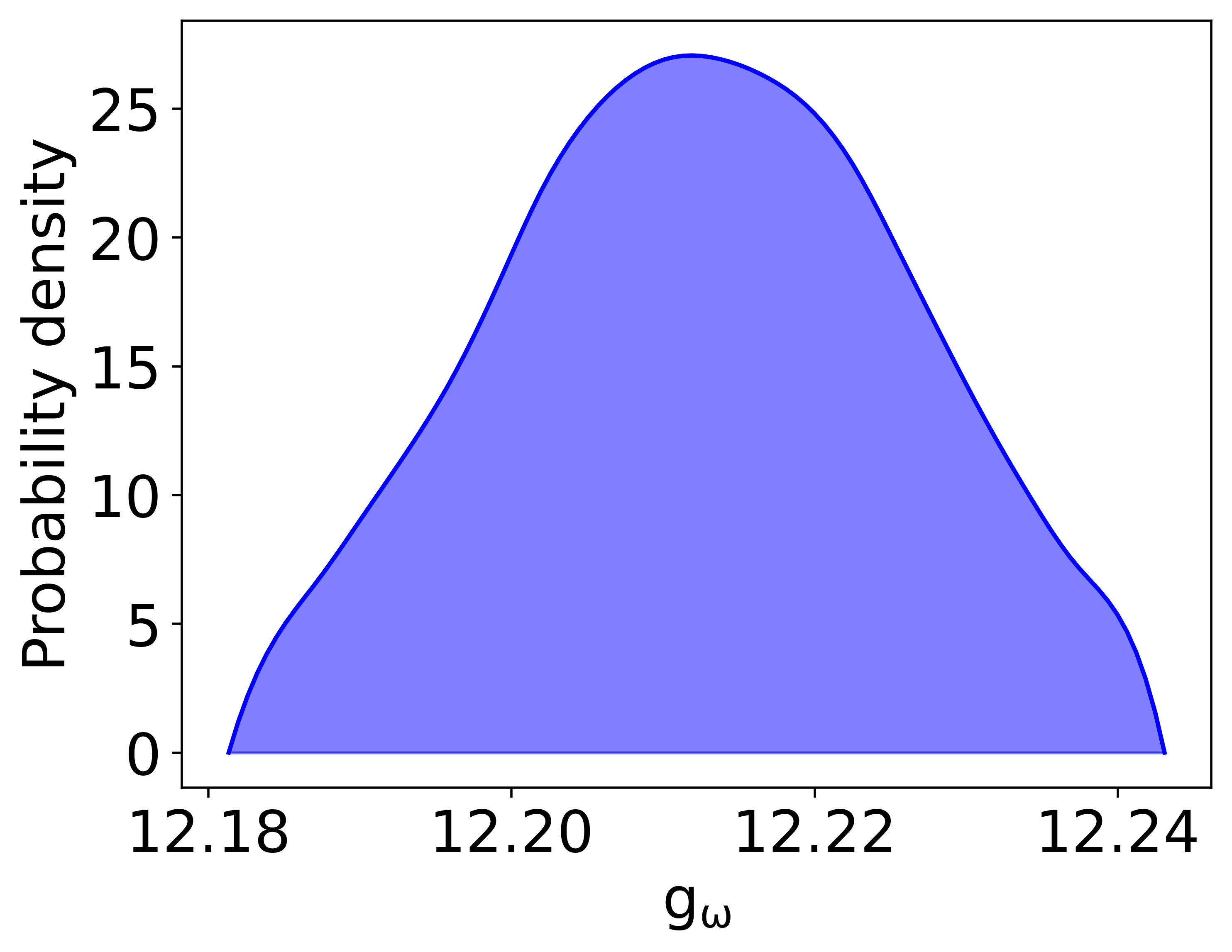}}
	\subfigure{\includegraphics[scale=0.25]{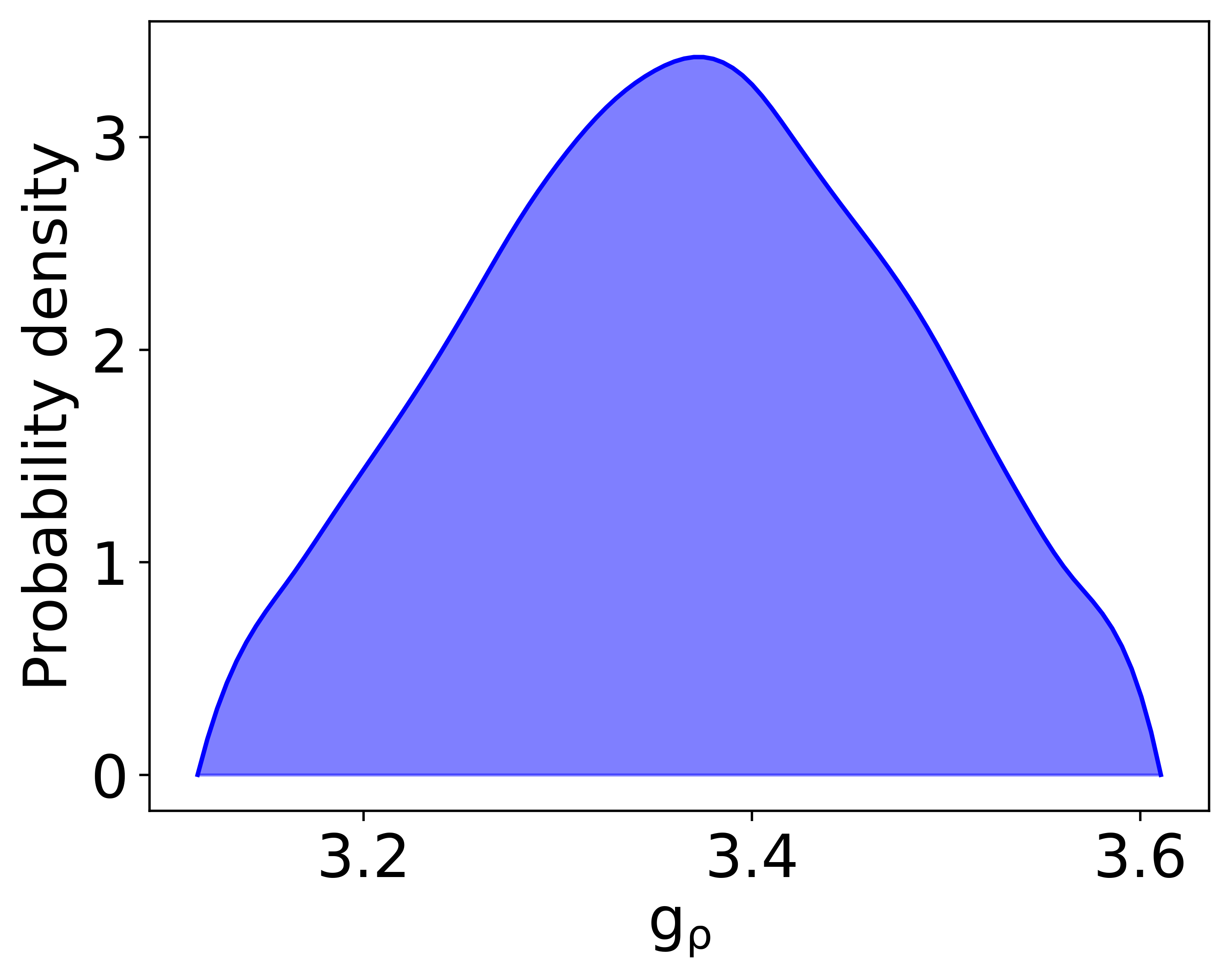}}
	\subfigure{\includegraphics[scale=0.25]{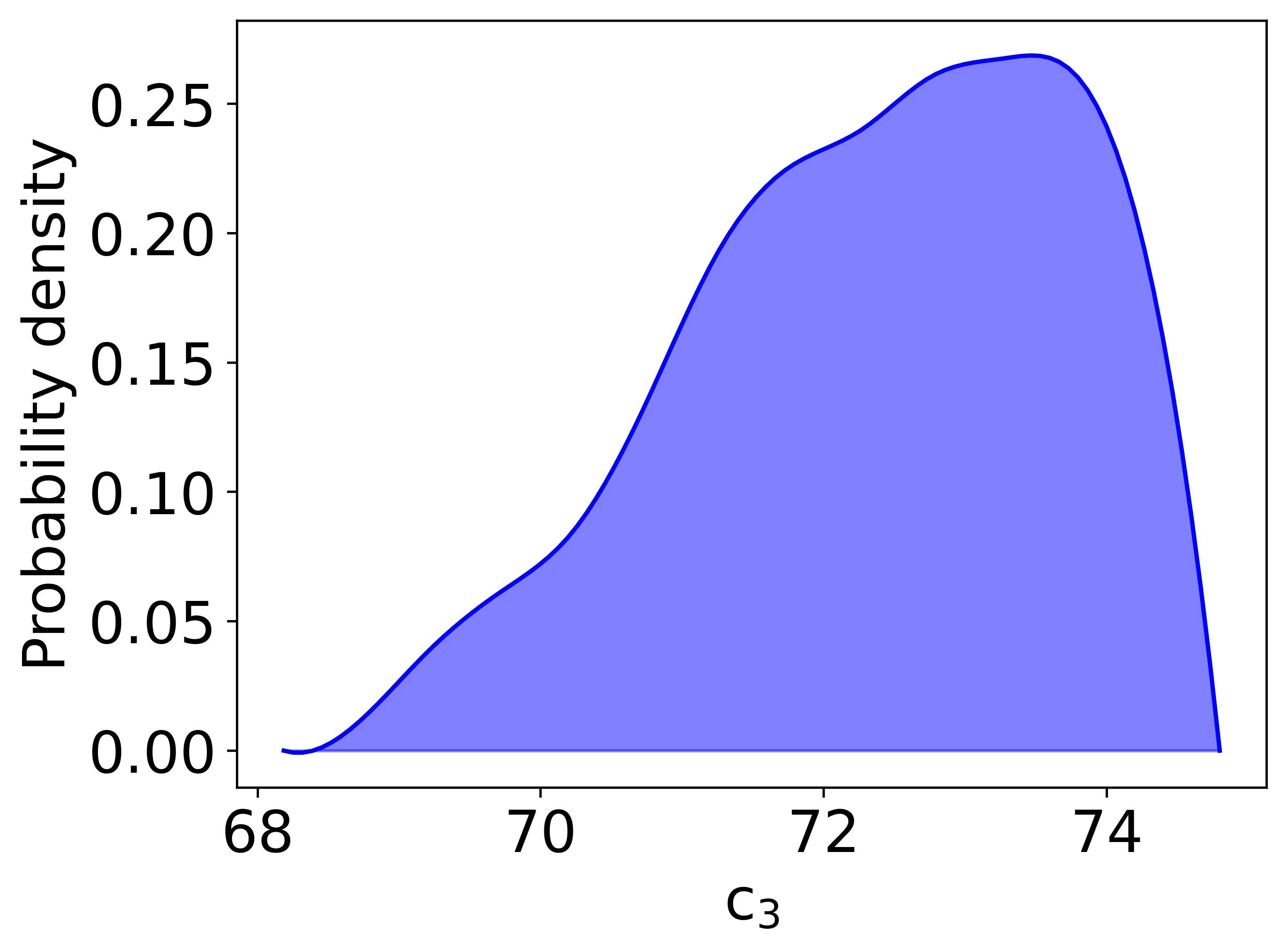}}
	\subfigure{\includegraphics[scale=0.25]{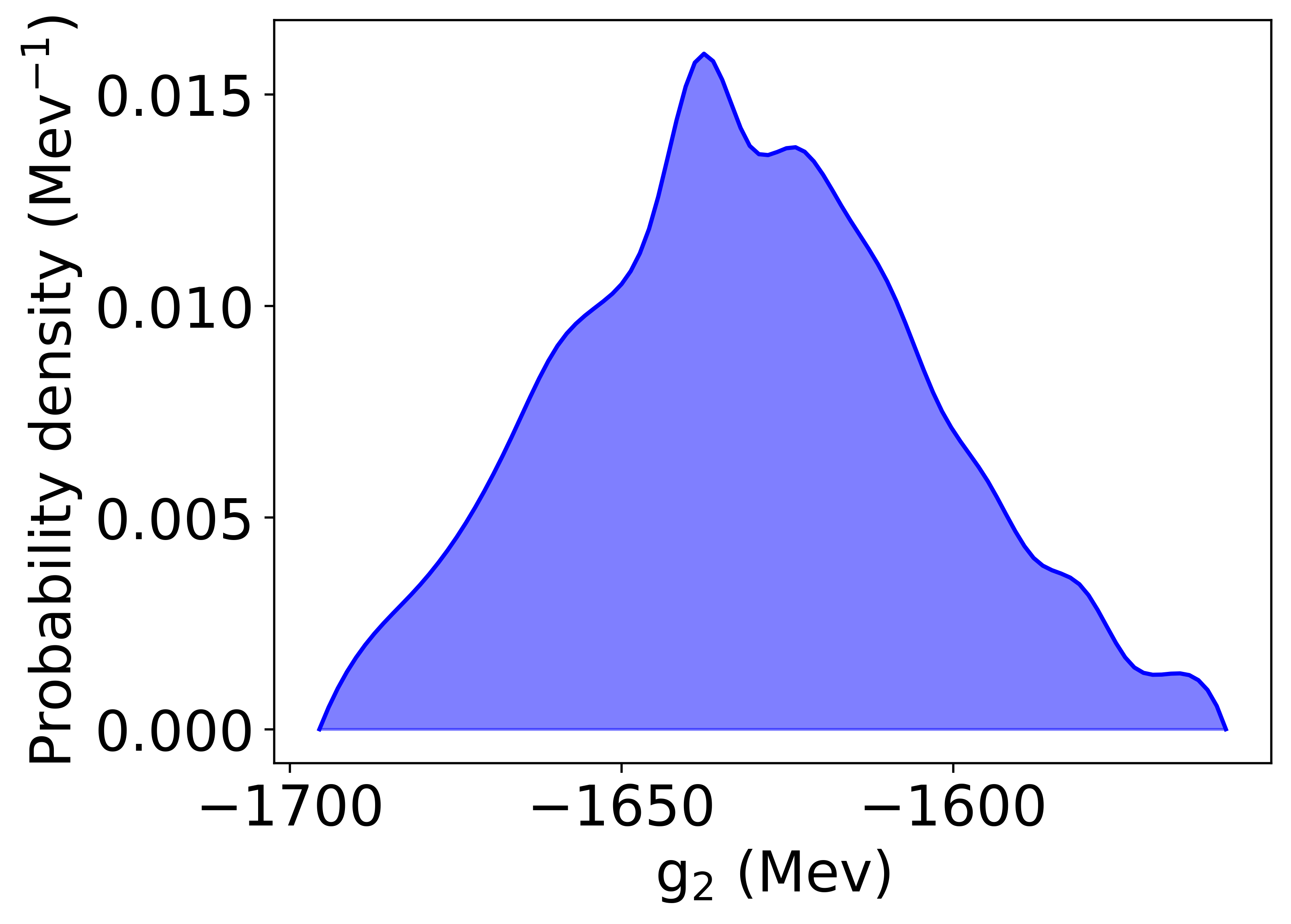}}
	\subfigure{\includegraphics[scale=0.25]{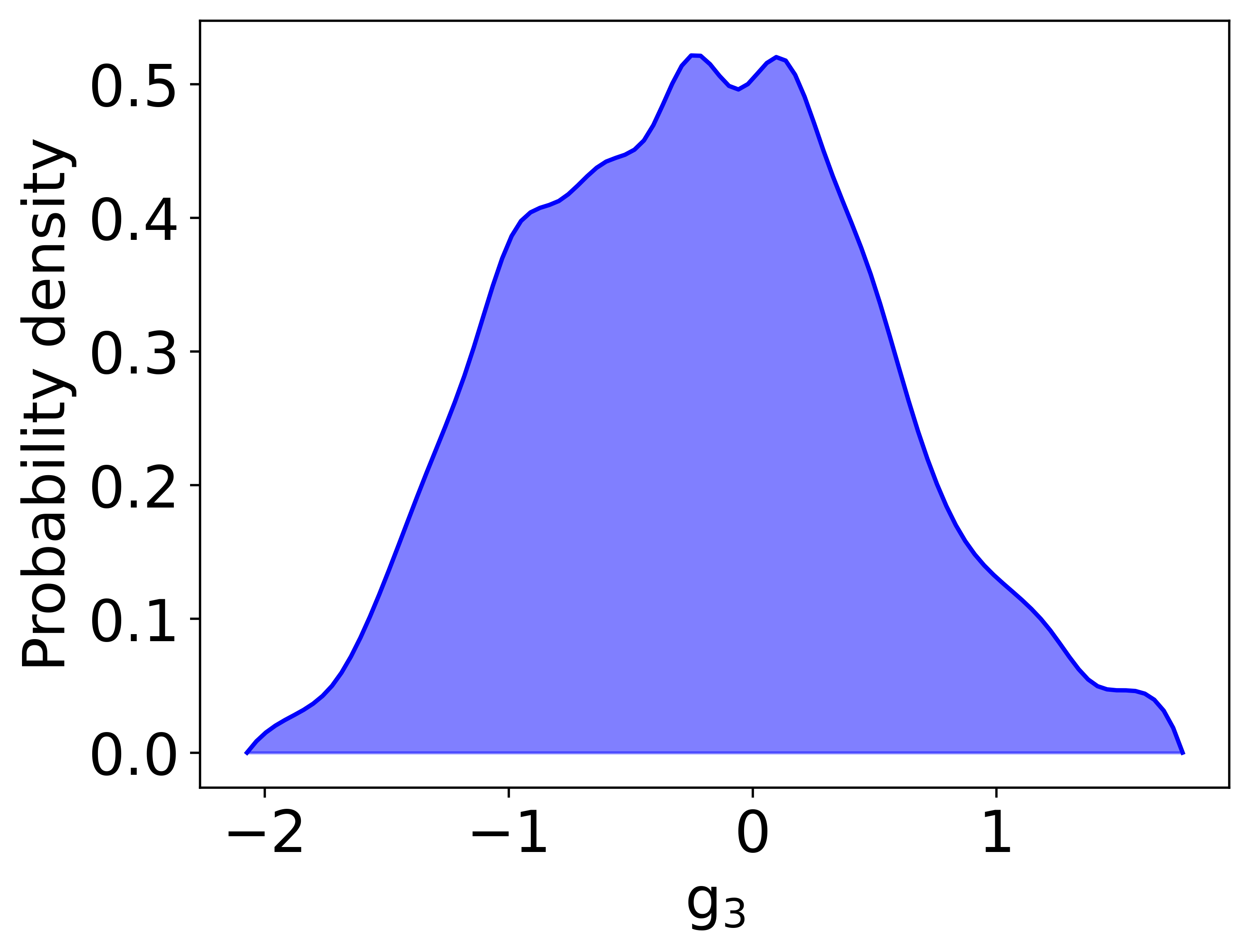}}
	\subfigure{\includegraphics[scale=0.25]{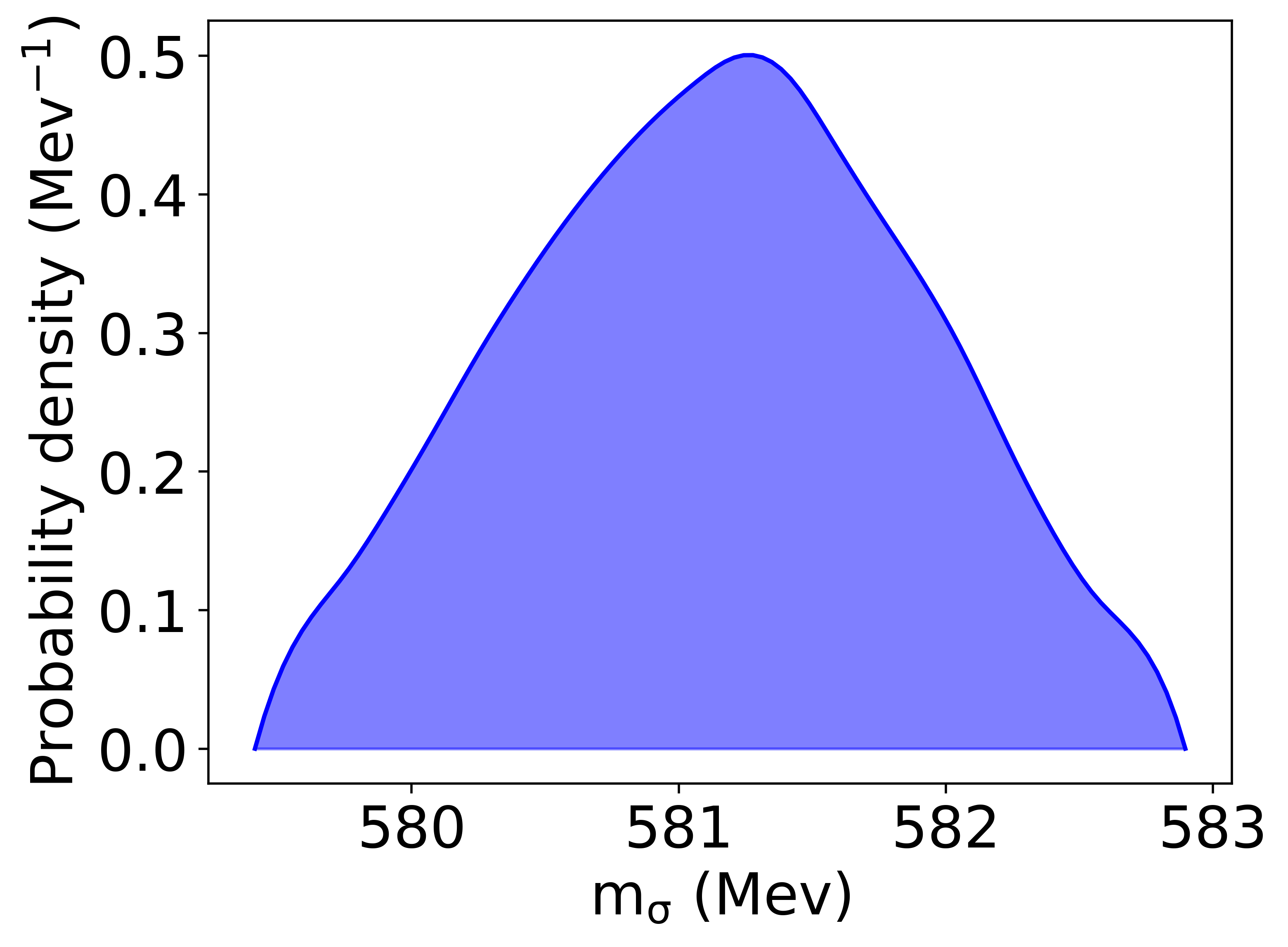}}
	\caption{Distributions of the parameters estimated by NN.}
	\label{fig:para-band}
\end{figure*}

Since parameter space is high dimensional, it is hard to fully describe the error bands of parameters. To have an idea of the distribution of a certain parameter, we vary it in the parameter space while fixing the others at the optimal values.~\footnote{The number of events is set to be 10000.} The distributions of the parameters are shown in Figure~\ref{fig:para-band}, and it can be seen that the probability density distributions of parameters are not regular normal distributions, which indicates the correlations between parameters, especially those multimeson couplings.

\section{Summary and Outlook}

An NN platform is constructed for analysis on bands of strong interaction parameters with respect to both NS properties and NM properties.
It has three main modules:
\begin{enumerate}
	\item The computation module is used to solve the coupled EoMs to obtain the corresponding physical quantities, and the LOSS function used during the training process is the EoM itself so that this kind of process can be applied to other physical problems straightforwardly.
	
	\item The Bayesian module is used to calculate the confidence level of the parameter space given by the computation module with reasonable physical error bars.
	
	\item The self-supervised module, the most important part, gives the platform the ability to search parameter space automatically and adaptively with the advantages of NN and will try to find the optimal solutions of parameters.
\end{enumerate}

As an example, we apply the constructed NN to a specific nucleon force model to show that the platform is able to give a reasonable prediction of parameter space. The numerical results suggest the necessity to analyze NS properties with the NM properties simultaneously since the predicted parameter space with only NM properties usually makes it hard to describe the MR relations.
The results also indicate the NN platform's ability to identify data-favored theoretical parameterization with the help of the optimal solution's confidence level, $P_c$. The NN applied in this example is modularly designed so that the computation module can be adjusted to solve other RMF EoS and the other two modules can be applied directly to find the optimal parameter space for corresponding physical data.

In our future works, other NS and NM properties, such as tidal deformations, the symmetry energy density slope, and the skewness coefficient, will be added to the Bayesian module to provide more rigorous constraints on parameters, and the algorithms involved will be changed into AI-governed ones to improve efficiency and adaptivity. The correlation analysis on parameter space will be carried out to distinguish the leading-order physical contribution parameters from the ones causing noise. The existing algorithms built in this work will be applied to other $\chi {\mathrm{EFTs}}$, such as the setup in~\cite{Li:2016uzn}, which are nonlinear realizations of QCD symmetry. The constraints on these LECs of the $\chi {\mathrm{EFTs}}$ can be of great help to understanding nonperturbative properties of QCD, such as quark condensate and trace anomaly at low-energy and dense regions.

In modern physics, multisource data handling and error propagations are becoming more and more important to help one understand EFTs and provide rigorous constraints on possible new physics.
The statistical analysis built in this platform can be extended to other model verification processes, especially those with massive data or complex mappings between physical quantities and parameters.
It is hoped to provide a heuristic way for theoretical physics research.

{\it Acknowledgements}:	The work of Y.~L. M. is supported in part by the National Key R\&D Program of China under grant No. 2021YFC2202900 and the National Science Foundation of China (NSFC) under grant No. 11875147 and No. 12147103.

\bibliographystyle{aasjournal}
\bibliography{AIMeanFieldRef}

\end{document}